\documentclass{svproc}
\usepackage{url}
\usepackage{graphicx}
\usepackage{multicol}
\usepackage{hyperref}

\begin{document}
\mainmatter              
\title{The beta-Oslo method: \\ experimentally constrained ($n,\gamma$) reaction rates relevant to the $r$-process}
\titlerunning{The beta-Oslo method}  
%
\author{A.~C.~Larsen\inst{1} \and S.~N.~Liddick\inst{2,}\inst{3}
A.~Spyrou\inst{2,}\inst{4,}\inst{5} \and M.~Guttormsen\inst{1} \and F.~L.~Bello~Garrote\inst{1} \and J.~E.~Midtb{\o}\inst{1} \and T.~Renstr{\o}m\inst{1}}
\authorrunning{A.C. Larsen et al.} 
%
\tocauthor{A.~C.~Larsen, S.~N.~Liddick, A.~Spyrou, M.~Guttormsen, F.~L.~Bello Garrote, J.~E.~Midtb{\o}, T.~Renstr{\o}m}
\institute{Department of Physics, University of Oslo, Norway,\\
\email{a.c.larsen@fys.uio.no},\\ 
\and
National Superconducting Cyclotron Laboratory, Michigan State University, USA,
\and
Department of Chemistry, Michigan State University, USA,
\and
Department of Physics and Astronomy, Michigan State University, USA,
\and
Joint Institute for Nuclear Astrophysics, Michigan State University, USA}

\maketitle              

\begin{abstract}
Unknown neutron-capture reaction rates remain a significant source of uncertainty 
in state-of-the-art $r$-process nucleosynthesis reaction network calculations. 
As the $r$-process involves highly neutron-rich nuclei for which direct ($n,\gamma$) 
cross-section measurements are virtually impossible, indirect methods are called for  
to constrain ($n,\gamma$) cross sections used as input for the $r$-process nuclear network. 
Here we discuss the newly developed beta-Oslo method, which is capable of provding experimental 
input for calculating ($n,\gamma$) rates of neutron-rich nuclei. 
The beta-Oslo method represents a first step towards constraining neutron-capture 
rates of importance to the $r$-process. 

\end{abstract}
%
\section{Introduction}
On August 17, 2017, the LIGO and Virgo gravitational-wave detectors measured, 
for the first time, a direct signal from two colliding neutron stars~\cite{abbott2017}. 
Follow-up measurements with telescopes sensitive to electromagnetic radiation confirmed 
that the rapid neutron-capture process ($r$-process)~\cite{burbidge1957,cameron1957} 
had indeed taken place in the collision (\textit{e.g.}, Ref.~\cite{Pian2017}). 
Hence, a long-standing question in nuclear astrophysics was at least partly solved; one 
astrophysical $r$-process site is now confirmed. 

However, the uncertain nuclear-physics input remains a huge obstacle in modeling the $r$-process
yields in large-scale nucleosynthesis network calculations~\cite{arnould2007,mumpower2016}. 
The $r$-process involves highly neutron-rich nuclei, where there is a severe lack of
relevant nuclear data such as masses, $\beta$-decay rates and neutron-capture cross sections. 
As shown in, \textit{e.g.}, Ref.~\cite{arnould2007}, one cannot rely on the 
assumption of ($n,\gamma$)--($\gamma,n$) equilibrium for typical $r$-process temperatures
and neutron densities in a neutron-star merger event, 
at least not at all times and for all trajectories 
as demonstrated in Ref.~\cite{mendoza-temis2015}.
As a consequence, neutron-capture rates will impact the final abundances 
and must be included in the nucleosynthesis calculations.
Moreover, it is an unfortunate fact that
different theoretical predictions for neutron-capture rates
may vary by several orders of magnitude.  

In this work, a recently developed method to address this issue is presented: 
The \textit{beta-Oslo method}~\cite{spyrou2014,liddick2016} provides data on the 
nuclear level density and average $\gamma$-decay strength of moderately neutron-rich nuclei. 
These quantities are crucial input for calculations of neutron-capture rates~\cite{arnould2007}. 
The beta-Oslo method presents a first step towards constraining neutron-capture rates 
of importance to the $r$-process.

\section{The Oslo and beta-Oslo methods}
The principles behind the beta-Oslo method are very similar to those of the Oslo method,
which will be briefly outlined in the following. 
The starting point is a set of excitation-energy tagged $\gamma$-ray spectra containing 
$\gamma$ rays from all possible cascades originating at a given initial excitation energy. 
In the Oslo method, this has been achieved by charged-particle--$\gamma$-ray 
coincidence measurements. 
The $\gamma$-ray spectra are corrected for the NaI 
detector response using the method described in Ref.~\cite{guttormsen1996}, 
and the distribution of primary $\gamma$ rays is determined
by an iterative subtraction technique~\cite{guttormsen1987}.
Finally, the nuclear level density (NLD) and 
$\gamma$-ray strength function ($\gamma$SF) are simultaneously 
extracted from the primary $\gamma$-ray distribution~\cite{schiller2000} 
and normalized to auxiliary data~\cite{larsen2011}.
The level-density and $\gamma$-strength data can then be used as input for ($n,\gamma$) 
cross-section calculations as shown, \textit{e.g.}, in Ref.~\cite{kheswa2015}.

In 2004, a surprising increase in the low-$\gamma$-energy region of the $\gamma$-decay strength
of $^{56,57}$Fe was discovered~\cite{voinov2004}. This \textit{upbend} has later been discovered
in many nuclei and has been confirmed with an independent measurement 
technique~\cite{wiedeking2012,jones2018} and shown to be dominantly of
dipole nature~\cite{larsen2013,simon2016}. If the upbend is indeed present in 
very neutron-rich nuclei such as those involved in the $r$-process, 
it could increase ($n,\gamma$) reaction rates by 1-2 orders of magnitude~\cite{larsen2010}.
Hence, it is critical to measure the $\gamma$SF in neutron-rich nuclei
to see whether the upbend exists in these exotic systems. 
To address this question and to provide indirect measurement
of ($n,\gamma$) reaction rates, 
the \textit{beta-Oslo method}~\cite{spyrou2014} was recently invented. 

The method exploits the high
$Q$-value for beta decay of neutron-rich nuclei, so that excited states
in a broad energy range will be populated in the daughter nucleus. 
Further, using a 
\textit{segmented} total-absorption spectrometer such as the 
SuN detector~\cite{SuN}, one obtains the sum of all $\gamma$ rays 
in the cascades giving the initial excitation energy,
while the single segments give the individual $\gamma$ rays. 
In this way, one can generate a matrix of excitation-energy tagged $\gamma$-ray spectra
and apply the Oslo method to extract NLD and $\gamma$SF for the
daughter nucleus. 

\begin{figure}[tb]
\begin{center}
\includegraphics[clip,width=1.0\columnwidth]{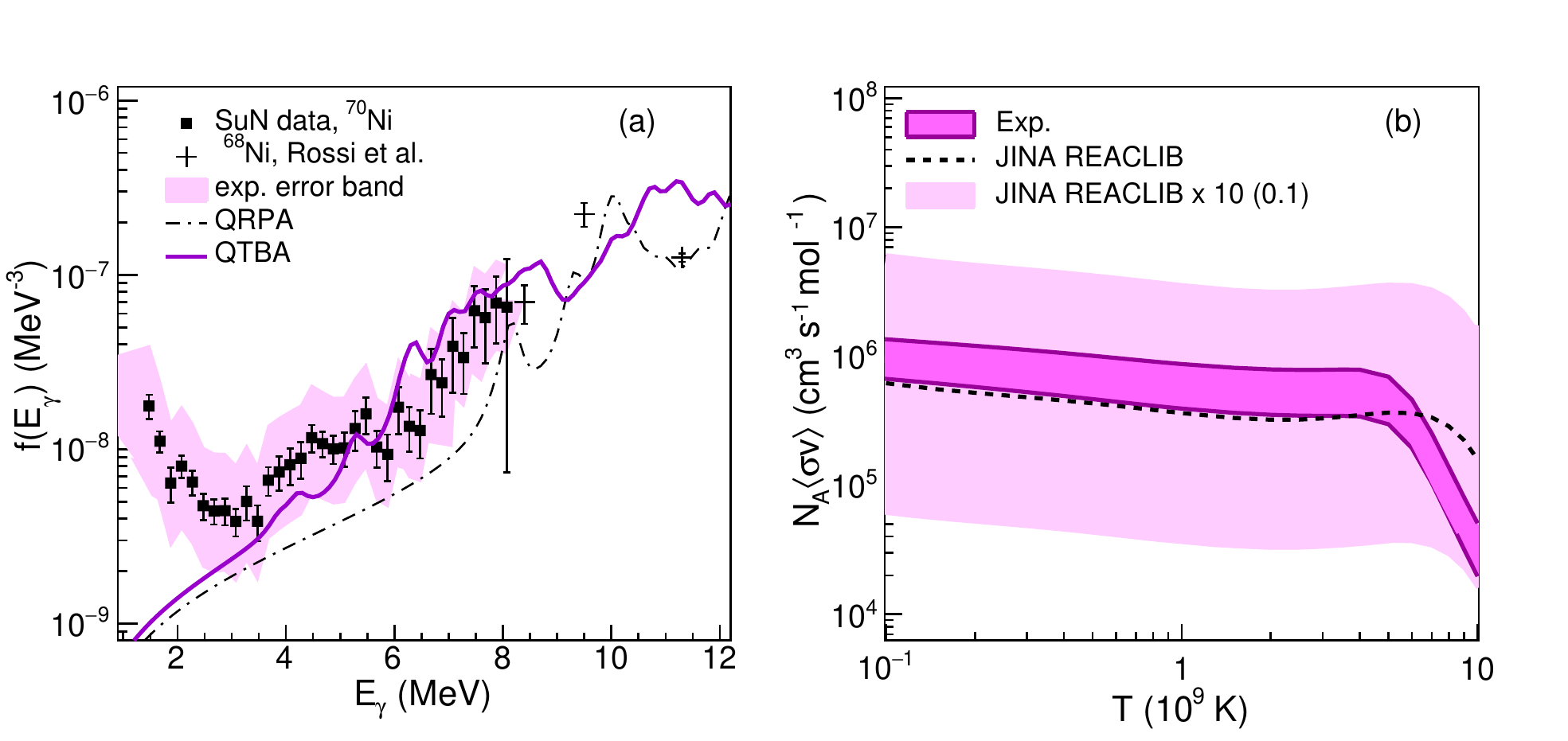}
\caption {(Color online) Gamma-decay strength function of $^{70}$Ni~\cite{larsen2018}
(a) and the $^{69}$Ni($n,\gamma$)$^{70}$Ni reaction rate from Ref.~\cite{liddick2016} (b), 
where we follow Ref.~\cite{surman2014} and compare with the
JINA REACLIB rate~\cite{JINA-REACLIB} (dashed line) scaled with a factor of 10 up and down
(light-shaded band).}
\label{fig:rates}
\end{center}
\end{figure}

The beta-Oslo method was first applied on $^{76}$Ga beta-decaying into 
$^{76}$Ge~\cite{spyrou2014}. The experiment was performed at the 
National Superconducting Cyclotron Laboratory (NSCL), Michigan State
University (MSU), using a 130-MeV/nucleon $^{76}$Ge beam producing 
$^{76}$Ga by fragmentation on a thick beryllium target. The $^{76}$Ga
secondary beam was 
implanted on an Si surface-barrier detector mounted inside SuN, 
which was used to measure the subsequent $\gamma$-ray cascades in the 
daughter nucleus $^{76}$Ge. The resulting data set enabled a 
significant improvement on the $^{75}$Ge($n,\gamma$)$^{76}$Ge reaction rate,
which has not been measured directly and so relied on purely theoretical 
estimates. 

Further, the beta-Oslo method has recently been applied on the  
neutron-rich $^{70}$Co isotope, beta-decaying into $^{70}$Ni~\cite{liddick2016}.
The experiment was performed at NSCL, MSU,
where a primary 140-MeV/nucleon $^{86}$Kr beam hit a beryllium target
to produce $^{70}$Co that was delivered to the experimental
setup, this time with a double-sided Si strip detector inside SuN. 
Again, SuN was used to detect the $\gamma$-ray cascades 
from the daughter nucleus, $^{70}$Ni. 
Complementary data from GSI on the $^{68}$Ni $\gamma$SF~\cite{rossi2013}
\textit{above} the neutron separation energy allowed for a well-determined
absolute normalization
of the full $\gamma$SF as shown in Fig.~\ref{fig:rates}a. 
The low-energy upbend is indeed present in the $^{70}$Ni $\gamma$SF and 
is likely due to strong low-energy $M1$ transitions as supported by 
shell-model calculations~\cite{larsen2018}. 

From the $^{70}$Ni data, the 
$^{69}$Ni($n,\gamma$)$^{70}$Ni reaction rate is deduced with an uncertainty  
of a factor $\sim 2-3$ (see Fig.~\ref{fig:rates}b and Ref.~\cite{liddick2016}). 
This is to be compared with the uncertainty band considered in Ref.~\cite{surman2014},
multiplying the JINA REACLIB rate~\cite{JINA-REACLIB} 
with a factor 0.1 and 10. It is clear that the data-constrained rate 
represents a significant improvement. 

\section{Summary and outlook}
The beta-Oslo method is capable of extracting NLDs and $\gamma$SFs of neutron-rich
nuclei, enabling an indirect way to experimentally constrain ($n,\gamma$) reaction 
rates of relevance to the $r$-process. 
So far, three reaction rates have been inferred: $^{75}$Ge($n,\gamma$)$^{76}$Ge~\cite{spyrou2014},
$^{69}$Ni($n,\gamma$)$^{70}$Ni~\cite{liddick2016} and 
$^{68}$Ni($n,\gamma$)$^{69}$Ni~\cite{spyrou2017}. 
In the future, many more rates will be constrained with this technique, to the benefit of
$r$-process nucleosynthesis calculations and our understanding of NLDs and $\gamma$SFs. 

\section*{Acknowledgments}
A.~C.~L. gratefully acknowledges funding 
through ERC-STG-2014 under grant agreement no. 637686. 
Support from the “ChETEC” COST Action (CA16117), 
supported by COST (European Cooperation in Science and Technology) is acknowledged.
This work was supported by the National Science Foundation under Grants No. PHY 1102511 (NSCL) 
and No. PHY 1430152 (JINA Center for the Evolution of the Elements), and PHY 1350234 (CAREER).
This material is based upon work supported by the US Department of Energy National Nuclear Security Administration through under Award No. DE-NA0003180, No. DE-NA-0000979 and No. DE-NA-0003221. 
%

%
%

\end{document}